\documentclass[%
 reprint,
 amsmath,amssymb,
 aps,
]{revtex4-2}

\usepackage{graphicx}
\usepackage{dcolumn}
\usepackage{bm}
\usepackage{color}
\usepackage{hyperref}
\hypersetup{
    colorlinks=true,
    linkcolor=blue,
    citecolor=blue,
}

\newcommand\mnras{MNRAS}             
\newcommand\aap{A\&A}                
\newcommand\apjl{ApJL}                
\newcommand\araa{ARA\&A}             
\newcommand\pasa{Publ. Astron. Soc. Australia}  
\newcommand\apjs{ApJS}

\begin{document}

\preprint{APS/123-QED}

\title{Upper limits on Einstein's weak equivalence principle placed by uncertainties of dispersion measures of fast radio bursts}

\author{Tetsuya Hashimoto}
\altaffiliation[Also at ]{Centre for Informatics and Computation in Astronomy (CICA), National Tsing Hua University, 101, Section 2. Kuang-Fu Road, Hsinchu, 30013, Taiwan (R.O.C.)}
\altaffiliation[Also at ]{Institute of Astronomy, National Tsing Hua University,\\101, Section 2. Kuang-Fu Road, Hsinchu, 30013, Taiwan (R.O.C.)}
\affiliation{Department of Physics, National Chung Hsing University, No. 145, Xingda Rd., South Dist., Taichung, 40227, Taiwan (R.O.C.)}
\email{tetsuya@phys.nchu.edu.tw}
\author{Tomotsugu Goto}
\affiliation{Institute of Astronomy, National Tsing Hua University,\\101, Section 2. Kuang-Fu Road, Hsinchu, 30013, Taiwan (R.O.C.)}
\author{Daryl Joe D. Santos}
\altaffiliation[Also at ]{Max Planck Institute for Extraterrestrial Physics, Gie{\ss}enbachstra{\ss}e 1, 85748 Garching, Germany}
\affiliation{Institute of Astronomy, National Tsing Hua University,\\101, Section 2. Kuang-Fu Road, Hsinchu, 30013, Taiwan (R.O.C.)}
\author{Simon C.-C. Ho}
\affiliation{Institute of Astronomy, National Tsing Hua University,\\101, Section 2. Kuang-Fu Road, Hsinchu, 30013, Taiwan (R.O.C.)}
\author{Ece Kilerci-Eser}
\affiliation{Sabanc{\i} University, Faculty of Engineering and Natural Sciences, 34956, Istanbul, Turkey}
\author{Tiger Y.-Y. Hsiao}
\affiliation{Institute of Astronomy, National Tsing Hua University,\\101, Section 2. Kuang-Fu Road, Hsinchu, 30013, Taiwan (R.O.C.)}
\author{Yi Hang Valerie Wong}
\affiliation{Institute of Astronomy, National Tsing Hua University,\\101, Section 2. Kuang-Fu Road, Hsinchu, 30013, Taiwan (R.O.C.)}
\author{Alvina Y. L. On}
\altaffiliation[Also at ]{Centre for Informatics and Computation in Astronomy (CICA), National Tsing Hua University, 101, Section 2. Kuang-Fu Road, Hsinchu, 30013, Taiwan (R.O.C.)}
\altaffiliation[Also at ]{Mullard Space Science Laboratory, University College London, Holmbury St Mary, Surrey RH5 6NT, UK}
\affiliation{Institute of Astronomy, National Tsing Hua University,\\101, Section 2. Kuang-Fu Road, Hsinchu, 30013, Taiwan (R.O.C.)}
\author{Seong Jin Kim}
\affiliation{Institute of Astronomy, National Tsing Hua University,\\101, Section 2. Kuang-Fu Road, Hsinchu, 30013, Taiwan (R.O.C.)}
\author{Ting-Yi Lu}
\altaffiliation[Also at ]{DARK, Niels Bohr Institute, University of Copenhagen, Jagtvej 128, 2200 Copenhagen, Denmark}
\affiliation{Institute of Astronomy, National Tsing Hua University,\\101, Section 2. Kuang-Fu Road, Hsinchu, 30013, Taiwan (R.O.C.)}

\date{Accepted 17 November 2021. Received 26 October 2021; in original form 24 May 2021}

\begin{abstract}
Fast radio bursts (FRBs) are astronomical transients with millisecond timescales occurring at cosmological distances.
The observed time lag between different energies of each FRB is well described by the inverse-square law of the observed frequency, i.e., dispersion measure.
Therefore, FRBs provide one of the ideal laboratories to test Einstein's weak equivalence principle (WEP): the hypothetical time lag between photons with different energies under a gravitational potential.
If WEP is violated, such evidence should be exposed within the observational uncertainties of dispersion measures, unless the WEP violation also depends on the inverse-square of the observed frequency.
In this work, we constrain the difference of gamma parameters ($\Delta\gamma$) between photons with different energies using the observational uncertainties of FRB dispersion measures, where $\Delta\gamma=0$ for Einstein's general relativity.
Adopting the averaged \lq Shapiro time delay\rq\ for cosmological sources, FRB 121002 at $z=1.6\pm0.3$ and FRB 180817.J1533+42 at $z=1.0\pm0.2$ place the most stringent constraints of $\log\Delta\gamma<-20.8\pm0.1$ and $\log(\Delta\gamma/r_{E}) < -20.9\pm0.2$, respectively, where $r_{E}$ is the energy ratio between the photons.
The former is about three orders of magnitude lower than those of other astrophysical sources in previous works under the same formalization of the Shapiro time delay while the latter is comparable to the tightest constraint so far.
\end{abstract}
\maketitle

\section{Introduction}
\label{introduction}
Einstein's general relativity (GR) \citep[][]{Einstein1907} is the basis of modern astronomy and astrophysics \citep{Abbott2016short, Ferreira2019, Abbott2019short,Isi2019}. 
Thus, testing the validity of basic assumptions made in GR is significant.
One of the basic assumptions of GR is the so-called \lq Einstein's weak equivalence principle (WEP)\rq.
WEP states that any uncharged free-falling test particle will follow a trajectory, which is independent of its internal composition and structure \citep[e.g.,][]{Will2006,Will2014}.
Any possible deviation from WEP is characterized by a $\gamma$ parameter for each particle whereas $\gamma=1$ for GR.
Here $\gamma$ describes how much space curvature is produced by a unit test mass \citep{Misner1973}.
The test particle can be massless such as photons and gravitational waves (GWs) \citep[e.g.,][]{Wei2015}.
Under the WEP assumption, different types of messenger particles (e.g., photons and GWs) must follow the same \lq Shapiro time delay\rq\ \citep{Shapiro1964} as far as they travel through the same gravitational field. 
This is also the case for the same-type particles with different internal properties such as energies and polarization states. 
Here, the Shapiro time delay is the time delay of a particle caused by gravitational fields in its path.
Therefore, the differences in Shapiro time delays between different particles (or the same particles with different internal properties) have been used to test WEP \citep[e.g.,][]{Wei2021}.

Fast radio bursts (FRBs) are new astronomical transients that show sudden brightening at radio wavelengths \citep[e.g.,][]{Lorimer2007}.
The timescale of FRBs is $\sim$ 1 millisecond \citep[e.g.,][]{Petroff2016}, and most FRBs are extragalactic events  \citep[e.g.,][]{Thornton2013short,Hashimoto2020c} likely to be encountering huge gravitational potentials, e.g., the Laniakea supercluster \citep[e.g.,][]{Tully2014}.
Some FRBs occurred at cosmological distances of $z\gtrsim1$ \citep{Hashimoto2020c}, where $z$ is redshift.
Therefore, FRBs provide one of the ideal laboratories to test WEP through the Shapiro time delays \citep[e.g.,][]{Wei2015,Tingay2016,Nusser2016,Xing2019}.
In this paper, we present new upper limits on the difference of $\gamma$ between photons with different energies using FRBs.
Throughout this paper, we focus on constraints provided by photons, GWs, and neutrinos with different energies.

The structure of the paper is as follows:
we describe our approach to constrain the WEP violation in Section \ref{method}.
In Section \ref{data}, the data used in this work is described.
Results and discussions are provided in Section \ref{results} and \ref{discussions}, respectively, followed by conclusions in Section \ref{conclusions}.

Throughout this paper, we assume the {\it Planck15} cosmology \citep{Planck2016short} as a fiducial model, i.e., $\Lambda$ cold dark matter cosmology with ($\Omega_{m}$,$\Omega_{\Lambda}$,$\Omega_{b}$,$H_{0}$)=(0.307, 0.693, 0.0486, 67.7), unless otherwise mentioned.

\section{Method}
\label{method}
The observed time delays of FRBs between different energies ($\Delta t_{\rm obs}$) can be expressed in terms of the contributions from at least the following components \citep[e.g.,][]{Wang2020}:
\begin{equation}
\label{dtobs}
\Delta t_{\rm obs} = \Delta t_{\rm DM} + \Delta t_{\rm int} + \Delta t_{\rm spe} + \Delta t_{\rm LIV} + \Delta t_{\rm gra},
\end{equation}
where $\Delta t_{\rm DM}$ is due to the so-called \lq dispersion measure (DM)\rq: the change of the speed of light depending on frequency when the radio emission passes through ionized materials in the host galaxy, intergalactic space, and the Milky Way.
$\Delta t_{\rm int}$ is the intrinsic time delay originating from the FRB source.
$\Delta t_{\rm spe}$ is the time delay from special-relativistic effects in the case of photons with nonzero rest mass.
$\Delta t_{\rm LIV}$ is the time delay from Lorentz invariance violation. 
$\Delta t_{\rm gra}$ is the difference of Shapiro time delay ($t_{\rm gra}$) \citep{Shapiro1964} which is caused by gravitational fields in the path of photons.
According to literature \citep[e.g.,][]{Wei2015,Gao2015,Xing2019,Minazzoli2019,Wang2020}, the upper limit on the WEP violation is estimated based on the following arguments. 
The terms $\Delta t_{\rm spe}$ and $\Delta t_{\rm LIV}$ in Eq. \ref{dtobs} are negligible compared to the other terms \citep[e.g.,][]{Wei2015,Gao2015,Wang2020}.
Assuming that $\Delta t_{\rm int}>0$ \citep[e.g.,][]{Wei2015}, Eq. \ref{dtobs} is approximated as 
\begin{equation}
\label{dtobs2}
\Delta t_{\rm obs} - \Delta t_{\rm DM} > \Delta t_{\rm gra}.
\end{equation}

Conventionally, $t_{\rm gra}$ and $\Delta t_{\rm gra}$ are parameterized by $\gamma$, which uses an approximation of the Minkowski metric with additional linear perturbation \citep{Wei2015,Gao2015,Xing2019,Minazzoli2019,Wang2020}.
However, such approximation is well justified only for the local Universe but is not the case for sources at cosmological distances of $z\gtrsim1$ \citep{Minazzoli2019}.
For cosmological sources, $t_{\rm gra}$ and $\Delta t_{\rm gra}$ do not monotonically increase with increasing gravitational potential sources \citep{Minazzoli2019}.
Therefore, assuming one gravitational source (conventionally, either the Milky Way or Laniakea supercluster \citep[e.g.][]{Wei2015,Gao2015,Xing2019,Wang2020}) does not provide a lower limit on $\Delta t_{\rm gra}$ anymore in Eq. \ref{dtobs2}.
In this sense, all of the gravitational sources near the light path needs to be taken into account to derive $\Delta t_{\rm gra}$ for cosmological sources.
However, such analysis is not practical using observational data of galaxies and galaxy clusters because galaxy observations are incomplete especially at higher redshifts (e.g., $z \gtrsim1$). 
Therefore, we use a cosmological analytic solution of $t_{\rm gra}$ for the averaged matter distribution \citep{Minazzoli2019}.
The averaged Shapiro time delay, $t_{\rm gra,ave}$, consists of two terms:
\begin{equation}
\label{tgra_ave}
t_{\rm gra,ave}= t_{\Lambda} + t_{\rm matter},
\end{equation}
where 
\begin{equation}
\label{tlambda}
t_{\Lambda}=\frac{\Omega_{\Lambda}H_{0}^{2}}{12c^{3}}d_{\rm S}^{3}
\end{equation}
and
\begin{equation}
\label{tmatter}
t_{\rm matter}=-\frac{\Omega_{m}H_{0}^{2}}{6c^{3}}d_{\rm S}^{3},
\end{equation}
where $d_{\rm S}$ is the comoving distance to the cosmological source.
Eq. \ref{tmatter} is consistent with the Shapiro time delay calculated from observed galaxy clusters, at least, up to $\sim$ 400 Mpc ($z\sim0.1$) \citep{Minazzoli2019}.
We caution that some works on the theoretical ground might be still needed to be sure that one can safely use the model given in \citet{Minazzoli2019} for this purpose.
We leave such theoretical studies for future works because the main focus of this paper is to present the advantage of FRBs over other astrophysical sources under the same assumptions on the Shapiro time delay.

Because we focus on the time lag under gravity in this work, we assume that the $t_{\Lambda}$ term is canceled out when $\Delta t_{\rm gra}$ is derived from two photons with different energies.
Using the matter contribution term ($t_{\rm matter}$) and the $\gamma$ parameter, $\Delta t_{\rm gra}$ is expressed as
\begin{equation}
\label{Deltatgra}
\Delta t_{\rm gra} = (\gamma_{2}-\gamma_{1})\frac{\Omega_{m}H_{0}^{2}}{6c^{3}}d_{\rm S}^{3}.
\end{equation}
Here, $\gamma_{1}$ and $\gamma_{2}$ are the gamma parameters of photons 1 and 2, respectively.
Eqs. \ref{dtobs2} and \ref{Deltatgra} provide
\begin{equation}
\label{Deltagamma}
\Delta \gamma := \gamma_{2}-\gamma_{1} < (\Delta t_{\rm obs}-\Delta t_{\rm DM})\frac{6c^{3}}{\Omega_{m}H_{0}^{2}d_{\rm S}^{3}}.
\end{equation}

In the FRB case, $\Delta t_{\rm obs}$ is well described by $\Delta t_{\rm DM}$ with a dependency of $\nu_{\rm obs}^{-2}$ \citep[e.g.,][]{Petroff2016,Macquart2020}, where $\nu_{\rm obs}$ is the observed frequency.
If WEP is violated, this effect should appear within the uncertainties of DM$_{\rm obs}$ measurements ($\delta$DM$_{\rm obs}$), where DM$_{\rm obs}$ is the observed dispersion measure. 
We note that this argument holds unless the WEP violation has such $\nu_{\rm obs}^{-2}$ dependency. 
In case both the WEP violation and $\Delta t_{\rm DM}$ follow the same $\nu_{\rm obs}^{-2}$ law, the two effects are degenerate (i.e., indistinguishable), and may cause systematically higher or lower values of observed dispersion measures than that of cosmological predictions \citep[e.g.,][]{Ioka2003,Inoue2004}, due to the additional dispersion by the WEP violation.
However, no such systematic difference has been reported \citep[e.g.,][]{Macquart2020}, indicating no clear evidence of the $\nu_{\rm obs}^{-2}$ law for the WEP violation.

The time lag due to DM$_{\rm obs}$ is approximated as
\begin{equation}
\Delta t_{\rm DM} \simeq 4.15\left( \frac{\nu_{\rm obs}}{\rm 1~GHz} \right)^{-2}\frac{\rm DM_{obs}}{10^{3} {\rm ~pc~cm^{-3}}}~{\rm s}
\end{equation}
\citep[e.g.,][]{Ioka2003,Inoue2004}.
The uncertainty of $\Delta t_{\rm DM}$ ($\delta \Delta t_{\rm DM}$) is proportional to $\delta$DM$_{\rm obs}$:
\begin{equation}
\label{eqdeltaDM}
\delta \Delta t_{\rm DM} \simeq4.15\left( \frac{\nu_{\rm obs}}{\rm 1~GHz} \right)^{-2}\frac{\rm \delta DM_{obs}}{10^{3} {\rm ~pc~cm^{-3}}}~{\rm s}.
\end{equation}
For some bright FRBs, DM$_{\rm obs}$ are accurately measured with $\delta{\rm DM}_{\rm obs}\lesssim$0.01 pc cm$^{-3}$ and corresponding $\delta\Delta t_{\rm DM}$ \citep[e.g.,][]{CHIMEFRB2019short}.
The time lag due to the WEP violation should be within $\delta\Delta t_{\rm DM}$ as mentioned above.
Therefore, $\delta$DM$_{\rm obs}$ places an upper limit on $\Delta\gamma$.
In this work, we use $\delta\Delta t_{\rm DM}$ as the upper limit on $\Delta t_{\rm obs}-\Delta t_{\rm DM}$ in Eq. \ref{Deltagamma}:
\begin{equation}
\label{eq8}
\Delta t_{\rm obs} - \Delta t_{\rm DM} < \delta\Delta t_{\rm DM}.
\end{equation}


\section{Data}
\label{data}
\begin{figure}
    \includegraphics[width=\columnwidth]{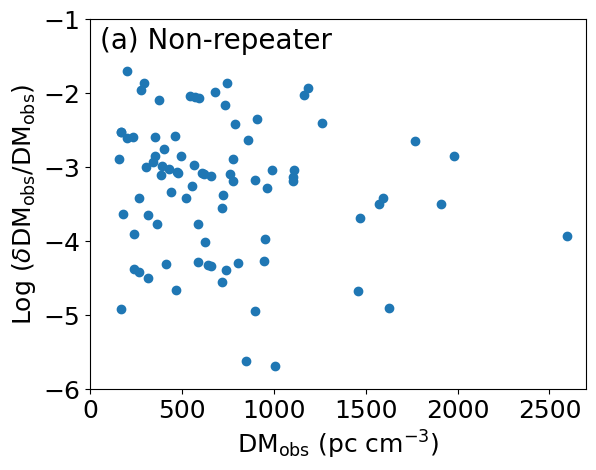}
    \includegraphics[width=\columnwidth]{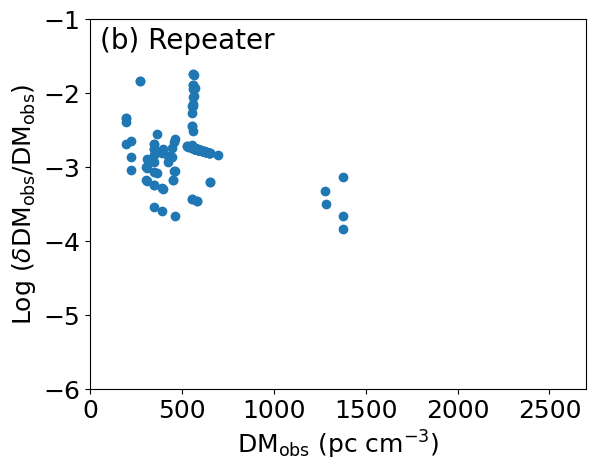}
    \caption{
    Log ($\delta$DM$_{\rm obs}$/DM$_{\rm obs}$) as a function of DM$_{\rm obs}$ for non-repeating FRBs (top) and repeating FRBs (bottom) in our sample. 
    For each repeating FRB source, multiple measurements of DM$_{\rm obs}$ and $\delta$DM$_{\rm obs}$ are shown due to the repetition.
    }
    \label{fig1}
\end{figure}

We use the FRB catalog \footnote{\url{http://www.phys.nthu.edu.tw/~tetsuya/archive/catalog/}} constructed by \citet{Hashimoto2020c}.
This catalog includes all the information from the FRBCAT project \citep{Petroff2016} as of 24 Feb. 2020 as well as complementary information on individual bursts of repeating FRBs compiled from literature \citep{Spitler2016short,Scholz2016short,Zhang2018,CHIMEFRB2019short,CHIME1repeat2019short,CHIME8repeat2019short,Kumar2019,Fonseca2020short}.
The catalog also includes redshifts of individual FRBs and their uncertainties calculated from DM$_{\rm obs}$ \citep[see][for details]{Hashimoto2020,Hashimoto2020c}.
In this work, we use DM$_{\rm obs}$, $\delta$DM$_{\rm obs}$, $\nu_{\rm obs}$, redshift, redshift uncertainty, and observed bandwidth of FRBs in the catalog.
The spectroscopic redshifts are utilized if they are available: FRB 121102, 180916.J0158+65, 180924, 181112, and 190523 \footnote{\url{http://frbhosts.org/}}.
Fig. \ref{fig1} shows $\delta$DM$_{\rm obs}$/DM$_{\rm obs}$ as a function of DM$_{\rm obs}$ for non-repeating and repeating FRBs.
Some non-repeating FRBs show $\log(\delta {\rm DM}_{\rm obs}/{\rm DM}_{\rm obs}) \sim -5$ which are one order of magnitude more accurate DM$_{\rm obs}$ measurements than those of repeating FRBs.
The mean values of $\log(\delta{\rm DM}_{\rm obs}/{\rm DM}_{\rm obs})$ of non-repeating and repeating FRBs are $-3.29\pm0.99$ and $-2.77\pm0.03$, respectively, where the uncertainties represent standard errors.
This is because the non-repeating FRBs are brighter than the repeating ones on average \citep[e.g.,][]{Hashimoto2020}.
According to Eqs. \ref{Deltagamma}, \ref{eqdeltaDM}, and \ref{eq8}, a more accurate DM$_{\rm obs}$ provides a stricter constraint on the time lag between different energies and thus $\Delta\gamma$.
We utilize both non-repeating and repeating FRBs in the following sections while non-repeating FRBs provide the most stringent constraints on $\Delta\gamma$ (see Section \ref{results}).

\section{Results}
\label{results}
\subsection{Tightest constraints in this work}
\label{tightest}
\begin{figure*}
    \includegraphics[width=2.0\columnwidth]{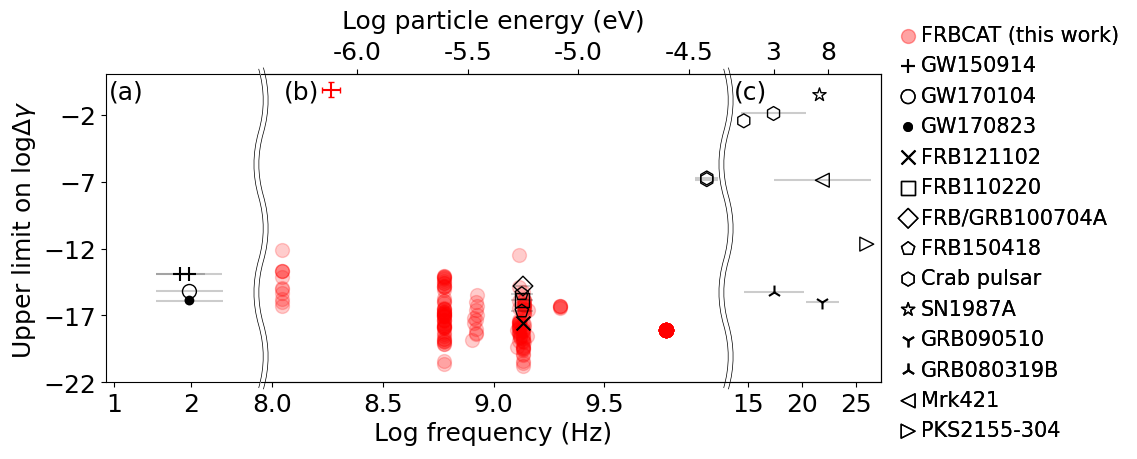}
    \caption{
    Upper limit on $\log \Delta\gamma$ as a function of observed frequency ($\nu_{\rm obs}$). 
    The same formalization of the Shapiro delay (Eq. \ref{Deltagamma}) is applied for both our sample and samples in previous works.
    (a) Limits constrained by gravitational wave (GW) sources, GW150914, GW170104, and GW170823 \citep{Wu2016,Kahya2016,Yang2020}. 
    Note that the frequencies represent the frequencies of GW signals.
    (b) Limits constrained by photons in radio and (c) photons in the eV-to-TeV range or neutrinos for SN1987A.
    The limits constrained by this work are shown by red dots.
    Their typical uncertainty (0.57 dex described in Section \ref{results}) and typical frequency range (20\% difference in frequency) are indicated by the red vertical and red horizontal error bars, respectively, at the top left corner of the panel (b).
    The cross, square, diamond, and pentagon, indicate upper limits derived from FRBs \citep{Xing2019,Wei2015,Tingay2016,Nusser2016}.
    The hexagons and a star are from the Crab pulsar \citep{Yang2016,Zhang2017,Desai2018,Leung2018} and SN1987A \citep{Longo1988}, respectively.
    The forked cross symbols indicate upper limits constrained from Gamma-ray bursts \citep[GRBs;][]{Gao2015}.
    The leftward and rightward triangles are from Blazars, Mrk421 and PKS2155$-$304 \citep{Wei2016}, respectively.
    The frequency range of each data point is indicated by the gray horizontal error bar unless the frequency range is smaller than the marker size.
    }
    \label{fig2}
\end{figure*}

Fig. \ref{fig2} shows the upper limits on $\log\Delta\gamma$ calculated by Eqs. \ref{Deltagamma}, \ref{eqdeltaDM}, and \ref{eq8} as a function of observed frequency (red dots) along with constraints in previous works \citep{Longo1988,Gao2015,Wei2015,Wu2016,Kahya2016,Tingay2016,Nusser2016,Yang2016,Wei2016,Zhang2017,Desai2018,Leung2018,Xing2019,Yang2020}.
For a fair comparison, the upper limits on $\log\Delta\gamma$ in previous works are re-calculated based on Eq. \ref{Deltagamma} using redshifts (or distances) and delay times adopted in the literature.
Each data shown in Fig. \ref{fig2} is derived from the time lag between the same particles (any of photons, GWs, and neutrinos) with different energies.
We note that the frequencies of the left panel in Fig. \ref{fig2} indicate the frequencies of GW signals \citep{Wu2016,Kahya2016,Yang2020} while those in the middle and right panels are observed frequencies of photons \citep[e.g.,][]{Gao2015} except for the neutrinos for SN1987A \citep{Longo1988}.
The upper limits of our sample derived from Eqs. \ref{Deltagamma}, \ref{eqdeltaDM}, and \ref{eq8} (red dots) are distributed down to $\log\Delta\gamma\sim$ $-20$.

The redshift uncertainties contribute to the uncertainties of $\log\Delta\gamma$ via Eq. \ref{Deltagamma}.
The median value of such uncertainties in our sample is 0.57.
Different assumptions on the cosmology also affect $\log\Delta\gamma$.
To evaluate the typical uncertainty due to the assumed cosmology, two sets of cosmological parameters are utilized to derive $\log\Delta\gamma$.
One is {\it Planck15} cosmology \citep{Planck2016short} with ($\Omega_{m}$,$\Omega_{\Lambda}$,$\Omega_{b}$,$H_{0}$)=(0.307, 0.693, 0.0486, 67.7), which is a fiducial model in this work.
Another is {\it WMAP5} cosmology \citep{Komatsu2009} with ($\Omega_{m}$,$\Omega_{\Lambda}$,$\Omega_{b}$,$H_{0}$)=(0.277, 0.723, 0.0459, 70.2).
The median value of $\log\Delta\gamma_{WMAP5}-\log\Delta\gamma_{Planck15}$ in our sample is 0.05.
The quadrature sum of uncertainties due to the redshift error and cosmology is adopted in this work.
The typical error of $\log\Delta\gamma$ is shown by the red vertical error bar in Fig. \ref{fig2}.

The most stringent constraint on $\log\Delta\gamma$ in this work is $\log\Delta\gamma<$ $-20.8\pm0.1$ which is provided by FRB 121002 at $z=1.6\pm0.3$ with $\delta$DM$_{\rm obs}=0.02$ pc cm$^{-1}$ and $\nu_{\rm obs}=1.2$-1.5 GHz \citep{Petroff2016}.
The re-calculated most stringent constraint in the previous works is $\log\Delta\gamma < -17.56\pm 0.05$ using Eq. \ref{Deltagamma} and FRB 121102 at $z=0.1927\pm0.00008$ \citep{Tendulkar2017} with the delay time of 0.4 ms between 1.344 and 1.374 GHz \citep{Xing2019}.
Therefore, our constraint on $\log\Delta\gamma$ is about three orders of magnitude tighter than those from other astrophysical sources in the previous works.

For FRB cases, the fractional energy differences between 
photons are typically $\sim20$\% \citep[e.g.,][]{Wei2015,Tingay2016,Nusser2016,Xing2019}.
In contrast, the high-energy astrophysical sources such as gamma-ray bursts (GRBs) and the Crab pulsar allow a comparison with much larger energy differences, e.g., more than three orders of magnitude \citep[e.g.,][]{Gao2015,Zhang2017}.
The deviation from the WEP may be more obvious for photons with larger energy differences if $\gamma$ is energy dependent. 
Therefore, the same $\Delta\gamma$ values constrained from different energy ratios might indicate different meanings.
To take such different energy ratios into account, \citet{Tingay2016} introduced $\log(\Delta\gamma/r_{E})$ where $r_{E} := E_{\rm high}/E_{\rm low}$ is the ratio of particle energies and $E_{\rm high}$ ($E_{\rm low}$) is the higher (lower) energy. 

\begin{figure*}
    \includegraphics[width=2.0\columnwidth]{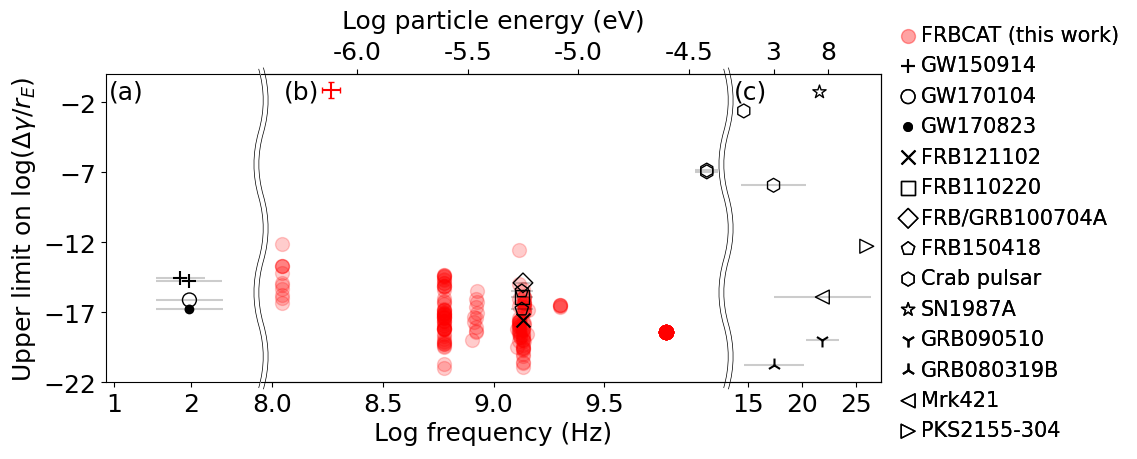}
    \caption{
    Same as Fig. \ref{fig2} except for $\log(\Delta\gamma/r_{E})$ in the vertical axis, where $r_{E} := E_{\rm high}/E_{\rm low}$ is the energy ratio between two particles with higher and lower energies ($E_{\rm high}$ and $E_{\rm low}$, respectively).
    For GW sources \citep{Wu2016,Kahya2016,Yang2020}, we assumed energies of gravitons which are proportional to their frequencies \citep[e.g.,][]{Modanese2004}.
    Adopted frequencies of two particles are 35-150 Hz \citep{Wu2016} and 35-250 Hz \citep{Kahya2016} for GW150914 and 35-256 Hz for GW170104 and GW170823 \citep{Yang2020}.
    }
    \label{fig3}
\end{figure*}

Fig. \ref{fig3} shows $\log(\Delta\gamma/r_{E})$ as a function of frequency, where $r_{E}$ in our sample is calculated from the observed bandwidth.
The upper limits on $\log(\Delta\gamma/r_{E})$ in previous works are re-calculated based on Eq. \ref{Deltagamma} using redshifts (or distances) and delay times adopted in the literature.
The most stringent constraint on $\log(\Delta\gamma/r_{E})$ in this work is $\log(\Delta\gamma/r_{E})<$ $-20.9\pm0.2$ which is provided by FRB 180817.J1533+42 at $z=1.0\pm0.2$ with $\delta$DM$_{\rm obs}=0.002$ pc cm$^{-1}$ and $\nu_{\rm obs}=0.4$-0.8 GHz \citep{CHIMEFRB2019short}.
The re-calculated most stringent constraint in the previous works is $\log(\Delta\gamma/r_{E}) < -20.77\pm 0.05$ using Eq. \ref{Deltagamma} and GRB 080319B at $z=0.937$ with the delay time of 5 s between 2 eV and 650 keV \citep{Gao2015}.
Our constraint on $\log(\Delta\gamma/r_{E})$ is comparable to the tightest constraint so far within the error.

\subsection{Robust constraints with spectroscopic redshifts}
In the FRBCAT catalog, there are five FRB sources with spectroscopic redshifts as of 18 Oct 2021: FRB 121102, 180916.J0158+65, 180924, 181112, and 190523 \citep{Tendulkar2017,Marcote2020,Bannister2019,Prochaska2019,Ravi2019host}.
Because these FRBs are almost free from the redshift uncertainties, the derived constraints on $\log\Delta\gamma$ are more robust than those provided by other FRBs.
For these FRBs, the uncertainty of $\log\Delta\gamma$ is dominated by the assumption on the cosmological parameters.
Therefore, we adopt 0.05 as the uncertainty of $\log\Delta\gamma$ due to the cosmology (see Section \ref{tightest} for details).
There are two repeating FRB sources out of five. 
Since such repeating FRB sources have multiple measurements of radio bursts, there are multiple measurements of $\log\Delta\gamma$ and $\log (\Delta\gamma/r_{E})$ for each repeating FRB source. 
We adopt the most stringent constraints for each FRB source. 
The derived robust constraints are summarized in Table \ref{tab1} together with the tightest constraints.

\begin{table*}
	\centering
	\caption{
	A summary of constraints on $\log\Delta\gamma$ and $\log(\Delta\gamma/r_{E})$ in this work.
	}
	\label{tab1}
	\begin{flushleft}
	\begin{tabular}{|l|c|c|c|c|}\hline
	\multicolumn{5}{|c|}{The tightest constraints} \\\hline
	FRB ID & Redshift & $\delta\Delta t_{\rm DM}$ & Adopted frequencies & Constraint \\
	       &          & (ms)       & (GHz)               &  \\\hline
	121002 & $1.6\pm0.3$ & 0.045     & 1.2-1.5             & $\log\Delta\gamma<-20.8\pm0.1$ \\
	180817.J1533+42 & $1.0\pm0.2$ & 0.023 & 0.4-0.8 & $\log(\Delta\gamma/r_{E})<-20.9\pm0.2$\\\hline
	\multicolumn{5}{|c|}{Robust constraints with spectroscopic redshifts} \\\hline
	121102$^{a}$ & 0.19273 & 0.115 & 4.0-8.0 & $\log\Delta\gamma<-18.10\pm0.05$, $\log(\Delta\gamma/r_{E})<-18.40\pm0.05$\\
	180916.J0158+65$^{a}$ & 0.0337 & 1.153 & 0.4-0.8 & $\log\Delta\gamma<-14.88\pm0.05$, $\log(\Delta\gamma/r_{E})<-15.18\pm0.05$\\
	180924 & 0.3214 & 0.143 & 1.2-1.5 & $\log\Delta\gamma<-18.63\pm0.05$, $\log(\Delta\gamma/r_{E})<-18.74\pm0.05$\\
	181112 & 0.4755 & 0.077 & 1.1-1.4 & $\log\Delta\gamma<-19.35\pm0.05$, $\log(\Delta\gamma/r_{E})<-19.47\pm0.05$\\
	190523 & 0.66 & 1.251 & 1.3-1.5 & $\log\Delta\gamma<-18.50\pm0.05$, $\log(\Delta\gamma/r_{E})<-18.55\pm0.05$\\\hline
    \end{tabular}\\
    $^{a}$ Since these are repeating FRB sources, the most stringent constraints are selected for each FRB source among values derived from multiple radio bursts.
    \end{flushleft}
\end{table*}


\section{Discussions}
\label{discussions}
FRBs have been used to constrain $\Delta\gamma$ in previous studies \citep{Wei2015,Tingay2016,Nusser2016,Xing2019}.
\citet{Wei2015} used $\Delta t_{\rm obs} (\sim 1~{\rm s})$ of FRB 110220 between 1.2 and 1.5 GHz.
They conservatively assumed that $\Delta t_{\rm obs}$ is dominated by the WEP violation rather than $\Delta t_{\rm DM}$ to obtain $\log\Delta\gamma < -7.6$.
\citet{Tingay2016} used $\Delta t_{\rm obs} (\sim 0.8~{\rm s})$ of FRB 150418 between 1.2 and 1.5 GHz.
Based on the same assumption by \citet{Wei2015}, \citet{Tingay2016} estimated $\log\Delta\gamma < -7.7$.
They argued that this limit would be reduced down to $\log\Delta\gamma < -9.0$ assuming that the WEP violation is masked by a $\sim 5$\% uncertainty of DM$_{\rm obs}$.
\citet{Nusser2016} proposed to use the gravitational potential of large-scale structures such as the Laniakea supercluster \citep[e.g.,][]{Tully2014} rather than the conventionally used Milky Way potential.
The potential fluctuations due to the large-scale structures can be used to constrain the Shapiro time delay and $\log \Delta\gamma$ at the cosmological scales \citep{Nusser2016} (see APPENDIX A for details of this approach using our sample).
The constraint estimated by \citet{Nusser2016} is $\log\Delta\gamma < -12$ to $-13$ for FRB 150418.
\citet{Xing2019} used sub-bursts of FRB 121102 at different frequencies of 1.344 and 1.374 GHz.
The time lag between these sub-bursts is $\Delta t_{\rm obs}=0.4$ ms, which provides $\log\Delta\gamma < -15.6$ assuming the Laniakea supercluster potential. 
These FRBs utilized for the WEP test are at $z<1$.
These previous works used an approximation of the Minkowski metric with additional linear perturbation.
\citet{Minazzoli2019} argued that such an approximation is not the case for cosmological sources.
Their revised formulae for cosmological sources (Eqs. \ref{tlambda} and \ref{tmatter}) allowed us to make use of distant FRBs at $z\gtrsim1$ to constrain $\Delta \gamma$ in this work.

Our approach is similar to the argument by \citet{Tingay2016} which takes $\delta$DM$_{\rm obs}$ into account.
They assumed 5\% as a typical uncertainty of DM$_{\rm obs}$.
However, we use exact values of $\delta$DM$_{\rm obs}$ for all of the extragalactic FRBs as of 24 Feb. 2020 \citep{Hashimoto2020c}, including FRBs at cosmological distances (e.g., $z\gtrsim$1).
In both $\log\Delta\gamma$ and $\log(\Delta\gamma/r_{E})$ cases, this work provides the most stringent constraints on WEP so far in the framework of the same particles with different energies.

In this work, we assumed that the $\Lambda$ terms of different particles (Eq. \ref{tlambda}) are canceled out because we focus on the time lag under gravity.
If this is not the case, the $\Lambda$ term has to be taken into account.
In such a case, the absolute values of $\log\Delta\gamma$ presented in this work would depend on the assumption on the $\Lambda$ term.
However, both the matter term and $\Lambda$ term show the same dependency on the source distance or redshift (Eqs. \ref{tlambda} and \ref{tmatter}).
Therefore, as far as the same formalization is utilized for different sources, the relative constraints on $\log\Delta\gamma$ do not change.
In this sense, this work still provides the most stringent constraints on the WEP violation so far even if the $\Lambda$ term is taken into account.


\section{Conclusion}
\label{conclusions}
FRBs are cosmological transients with millisecond timescales.
The observed time lag between different energies of each FRB is well described by the dispersion measure.
The time lag due to the dispersion measure follows the $\nu_{\rm obs}^{-2}$ law. 
Therefore, FRBs allow us to test the WEP violation, which is the hypothetical time lag between photons with different energies under a gravitational potential.
If WEP is violated, such evidence should appear within the observational uncertainties of dispersion measures unless the WEP violation also has the $\nu_{\rm obs}^{-2}$ dependency.

In this work, we test the time lag between photons with different energies using the observational uncertainties of dispersion measures of FRBs.
Adopting the analytic formula of the averaged Shapiro time delay for cosmological sources, the most stringent constraints are $\log\Delta\gamma<-20.8\pm0.1$ for FRB 121002 at $z=1.6\pm0.3$ and $\log(\Delta\gamma/r_{E}) < -20.9\pm0.2$ for FRB 180817.J1533+42 at $z=1.0\pm0.2$.
The former is about three orders of magnitude lower than those of other astrophysical sources in previous works, including GWs, FRB 121102, 110220, FRB/GRB100704A, 150418, the Crab pulsar, SN1987A, GRBs, and Blazars under the same formalization of the Shapiro time delay.
The latter is comparable to the tightest constraint so far.
Much larger number of FRBs are expected to be discovered in the near future.
Cosmological FRBs and the uncertainties of dispersion measures have a great potential to test the WEP violation accurately.


\section*{Acknowledgements}
We are very grateful to the anonymous referees for many insightful comments.
AYLO is supported by the Centre for Informatics and Computation in Astronomy (CICA) at National Tsing Hua University (NTHU) through a grant from the Ministry of Education of Taiwan.
TG and TH acknowledge the supports of the Ministry of Science and Technology of Taiwan through grants 108-2628-M-007-004-MY3 and 110-2112-M-005-013-MY3, respectively.
AYLO's visit to NTHU was supported by the Ministry of Science and Technology of the ROC (Taiwan) grant 105-2119-M-007-028-MY3, hosted by Prof. Albert Kong.
This work used high-performance computing facilities operated by the CICA at NTHU. 
This equipment was funded by the Ministry of Education of Taiwan, the Ministry of Science and Technology of Taiwan, and NTHU.
This research has made use of NASA's Astrophysics Data System.
The authors thank the Yukawa Institute for Theoretical Physics at Kyoto University. 
Discussions during the YITP International Molecule-type Workshop \lq Fast Radio Bursts: A Mystery Being Solved?\rq\ were useful to complete this work.
\section*{Data availability}
The data underlying this article is publicly available at FRBCAT project (\url{http://frbcat.org/}) and references therein.
The compiled catalog is available at \url{http://www.phys.nthu.edu.tw/~tetsuya/archive/catalog/}.

\appendix
\renewcommand{\thefigure}{A\arabic{figure}}
\setcounter{figure}{0}
\section{approach by Nusser 2016}
\begin{figure*}
    \includegraphics[width=2.0\columnwidth]{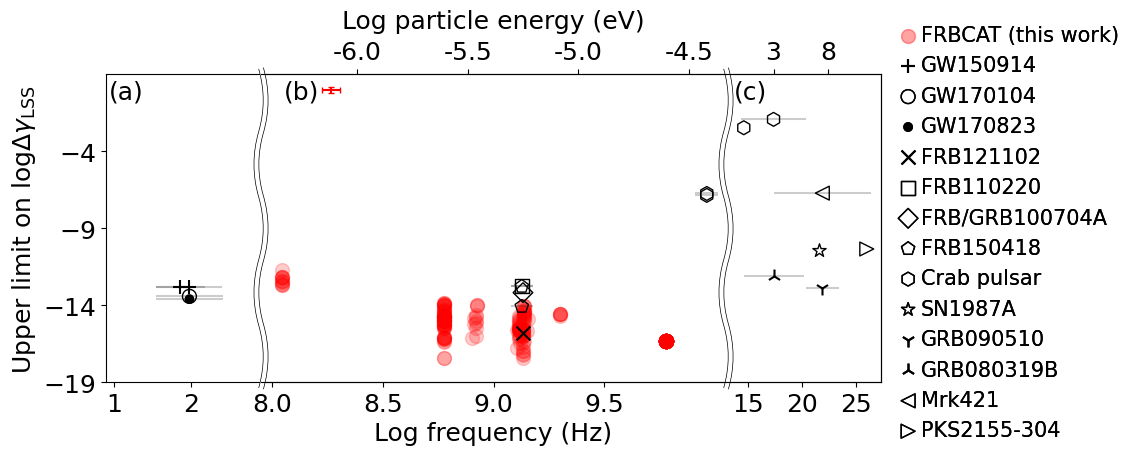}
    \caption{
    Same as Fig. \ref{fig2} except for the large-scale structure (LSS) approach by \citet{Nusser2016}.
    Note that $\log\Delta\gamma$ of the Crab pulsar are based on Eq. \ref{Deltagamma} because its redshift is not available due to a Galactic source.
    }
    \label{figA1}
\end{figure*}

\begin{figure*}
    \includegraphics[width=2.0\columnwidth]{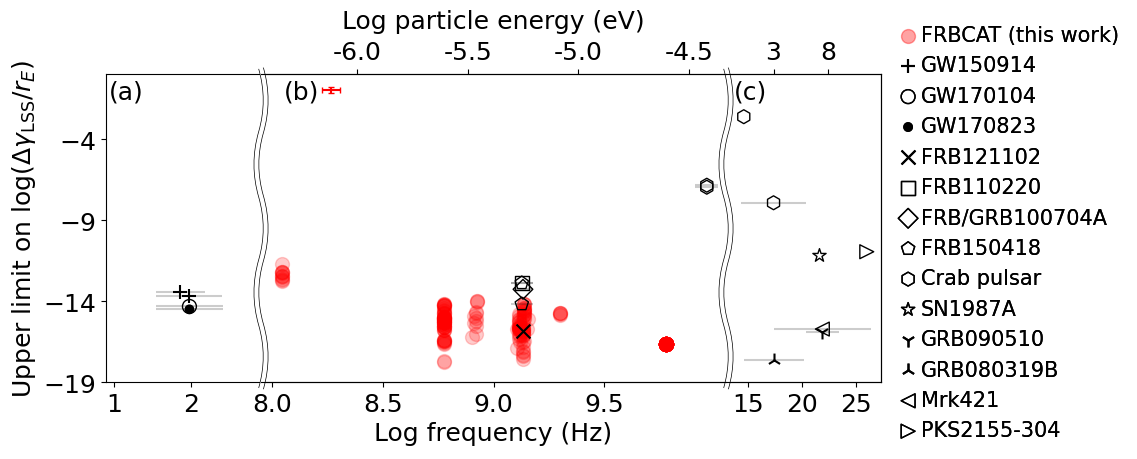}
    \caption{
    Same as Fig. \ref{fig3} except for the large-scale structure (LSS) approach by \citet{Nusser2016}.
    Note that $\log\Delta\gamma$ of the Crab pulsar are based on Eq. \ref{Deltagamma} because its redshift is not available due to a Galactic source.
    }
    \label{figA2}
\end{figure*}

\citet{Nusser2016} proposed to use potential fluctuations due to the large-scale structures (LSSs) to constrain the Shapiro time delay and $\log \Delta\gamma_{\rm LSS}$ at the cosmological scales.
The subscript \lq LSS\rq\ indicates the LSS approach by \citet{Nusser2016}.
The evaluation of the Shapiro delay in \citet{Nusser2016} is one of the only estimations of the Shapiro delay that has not been invalidated by the work of \citet{Minazzoli2019}.
To follow their approach, we fit a two-term exponential function to $\log\Delta\gamma_{\rm LSS} (t_{\rm gra}=1~{\rm s})$ as a function of redshift presented in Fig. 1 in \citet{Nusser2016}.
The best-fit function is
\begin{equation}
\label{Nusser}
\log \Delta\gamma_{\rm LSS}=-12.85+0.85e^{-z/0.09}+0.55e^{-z/0.43},
\end{equation}
where $\log \Delta\gamma$ is calculated for $t_{\rm gra}=1$ s.
We calculate $\log\Delta\gamma_{\rm LSS}$ for both our FRB sample and samples in the previous works shown in Figs. \ref{fig2} and \ref{fig3} using their redshifts, delay times, and Eq. \ref{Nusser}.
The results are shown in Figs. \ref{figA1} and \ref{figA2}.

In this approach, the most stringent constraint on $\log\Delta\gamma_{\rm LSS}$ in this work is $\log\Delta\gamma_{\rm LSS}<$ $-17.43\pm0.05$ which is provided by FRB 180817.J1533+42 at $z=1.0\pm0.2$ with $\delta$DM$_{\rm obs}=0.002$ pc cm$^{-1}$ and $\nu_{\rm obs}=0.4$-0.8 GHz \citep{CHIMEFRB2019short}.
The re-calculated most stringent constraint among the previous samples is $\log\Delta\gamma_{\rm LSS} < -15.80\pm 0.05$ derived from FRB 121102 at $z=0.1927\pm0.00008$ \citep{Tendulkar2017} with the delay time of 0.4 ms between 1.344 and 1.374 GHz \citep{Xing2019}.
Our constraint on $\log\Delta\gamma_{\rm LSS}$ is more than one order of magnitude tighter than those from other astrophysical sources in the previous works.

FRB180817.J1533+42 also provides the tightest constraint of $\log(\Delta\gamma_{\rm LSS}/r_{E}) < -17.73\pm0.05$ in this work.
This value is comparable to the tightest constraint among the previous samples, $\log(\Delta\gamma_{\rm LSS}/r_{E}) < -17.60\pm0.05$, derived from GRB 080319B at $z=0.937$ with the delay time of 5 s between 2 eV and 650 keV \citep{Gao2015}.

\newcommand{\noopsort}[1]{}
%

\end{document}